\documentclass[sigconf]{acmart}
\usepackage{amsmath}
\usepackage{color}
\usepackage{subfigure}
\usepackage{array}
\usepackage{{inputenc}}
\usepackage{balance}
\usepackage{multirow,tabularx}
\usepackage{booktabs,longtable}
\usepackage{algorithm}
\usepackage{algorithmic}
\usepackage{graphicx}
\usepackage{enumitem}
\usepackage{verbatim}
\usepackage{amsfonts}
\usepackage{hyperref}
\hypersetup{colorlinks=false,linkcolor=blue,urlcolor=blue,citecolor=red}

\newcommand{\Lapl}{\mathbf{\mathop{\mathcal{L}}}}

\newcommand{\Mat}[1]{\textbf{#1}}

\newcommand{\Space}[1]{\mathbb{#1}}
\newcommand{\Set}[1]{\mathcal{#1}}

\newcommand{\ie}{\emph{i.e., }}
\newcommand{\eg}{\emph{e.g., }}

\newcommand{\wrt}{\emph{w.r.t. }}
\newcommand{\cf}{\emph{cf. }}
\newcommand{\aka}{\emph{aka. }}

\AtBeginDocument{%
  \providecommand\BibTeX{{%
    \normalfont B\kern-0.5em{\scshape i\kern-0.25em b}\kern-0.8em\TeX}}}

\copyrightyear{2022}
\acmYear{2022}
\setcopyright{acmcopyright}
\acmConference[SIGIR '22] {Proceedings of the 45th International ACM SIGIR Conference on Research and Development in Information Retrieval}{July 11--15, 2022}{Madrid, Spain.}
\acmBooktitle {Proceedings of the 45th International ACM SIGIR Conference on Research and Development in Information Retrieval (SIGIR '22), July 11--15, 2022, Madrid, Spain}
\acmPrice{15.00}
\acmISBN{978-1-4503-8732-3/22/07}
\acmDOI{10.1145/XXXXXX.XXXXXX}

\begin{document}
\fancyhead{}

\title{HAKG: Hierarchy-Aware Knowledge Gated Network for Recommendation}

\author{Yuntao Du$^1$, Xinjun Zhu$^2$, Lu Chen$^1$, Baihua Zheng$^3$, and Yunjun Gao$^{1*}$}

\def \authors{Yuntao Du, Xinjun Zhu, Lu Chen, Baihua Zheng, and Yunjun Gao}

\affiliation{
 {\large$^1$}College of Computer Science, Zhejiang University, Hangzhou, China\\
 {\large$^2$}School of Software, Zhejiang University, Ningbo, China\\
 {\large$^3$}School of Computing and Information Systems, Singapore Management University, Singapore
 \country{}}

\email{{ytdu,xjzhu,luchen}@zju.edu.cn, bhzheng@smu.edu.sg, gaoyj@zju.edu.cn}

\thanks{$^*$Yunjun Gao is the corresponding author.}
\renewcommand{\shortauthors}{Du, et al.}
\begin{abstract}
Knowledge graph (KG) plays an increasingly important role to improve the recommendation performance and interpretability. A recent technical trend is to design end-to-end models based on information propagation schemes. However, existing propagation-based methods fail to (1) model the underlying hierarchical structures and relations, and (2) capture the high-order collaborative signals of items for learning high-quality user and item representations.

In this paper, we propose a new model, called \textit{Hierarchy-Aware Knowledge Gated Network} (HAKG), to tackle the aforementioned problems. Technically, we model users and items (that are captured by a user-item graph), as well as entities and relations (that are captured in a KG) in hyperbolic space, and design a hyperbolic aggregation scheme to gather relational contexts over KG. Meanwhile, we introduce a novel angle constraint to preserve characteristics of items in the embedding space. Furthermore, we propose a dual item embeddings design to represent and propagate collaborative signals and knowledge associations separately, and leverage the gated aggregation to distill discriminative information for better capturing user behavior patterns. Experimental results on three benchmark datasets show that, HAKG achieves significant improvement over the state-of-the-art methods like CKAN, Hyper-Know, and KGIN. Further analyses on the learned hyperbolic embeddings confirm that HAKG offers meaningful insights into the hierarchies of data. 
\end{abstract}

\ccsdesc[500]{Information systems~Recommender systems}

\keywords{Recommendation, Graph Neural Network, Knowledge Graph}

\maketitle

\section{Introduction}
\label{sec:introduction}

\begin{figure}[t]
	\centering
	\subfigure[Knowledge-aware recommendation]{
		\includegraphics[width=0.220\textwidth]{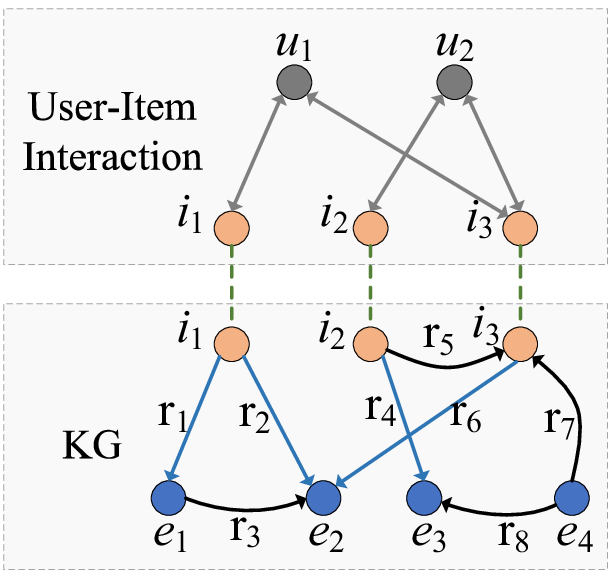}
		\label{fig:kg}}
	\hspace{0.02mm}
	\subfigure[Distribution of interactions]{
		\includegraphics[width=0.20\textwidth]{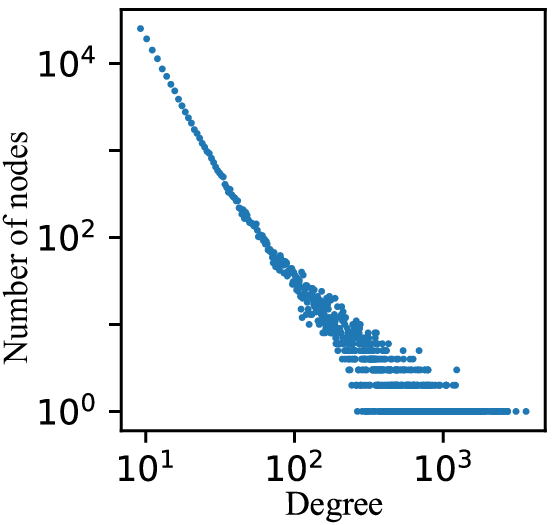}
		\label{fig:distribution_alibaba}}
	\vspace{-10pt}
	\caption{(a) An example of knowledge-aware recommendation. The blue lines indicate the hierarchical relations, while the black lines denote the non-hierarchical relations. (b) Degree distribution of Alibaba-iFashion dataset.
	}
	\label{fig:distribution and KG4Rec}
	\vspace{-15pt}
\end{figure}

In the era of information explosion, recommender systems have become an essential part of Internet applications to provide personalized information services. Traditional recommender systems that are based on collaborative filtering ~\cite{kdd08cf,icdm08cf,sigir19ngcf,gharibshah2020deep} usually suffer from data sparsity and cold-start problems. Recently, knowledge graph (KG), which provides various real-world facts related to items via relations, has demonstrated an impressive ability to alleviate cold-start issue and improve the explainability of recommendations.

Learning user and item representations from auxiliary KG has become the term of knowledge-aware recommendation. Early studies~\cite{kdd16cke,www18dkn,alg18cfkg} directly integrate knowledge graph embeddings with items to enhance their representations. Some subsequent studies~\cite{recsys16path,kdd18metapath,aaai19explainable,kdd20hinrec}
enrich the interactions via meta-paths from users to items for better identifying user-item connectivities. However, in order to obtain informative paths, these methods suffer from labor-intensive process~\cite{aaai19explainable}, poor generalization~\cite{kdd18metapath,recsys18rkge}, and unstable performance~\cite{sigir19reinforceKG}. Recently, the success of Graph Neural Networks (GNNs)~\cite{iclr17gcn,nips17sageGCN,iclr18gat} has inspired the community to develop end-to-end models based on the information aggregation schemes~\cite{cikm18ripple,kdd19kgat,kdd19kgnn-ls,sigir20ckan,www21kgin,cikm21kcan,wsdm22hyperRS,tkde22metakg}. The key idea is to iteratively propagate high-order information over KG, which could effectively integrate multi-hop neighbors into representations and hence improve the recommendation performance.

Although existing propagation-based methods are able to achieve good performance, we would like to highlight that they all fail to model the following two important factors.
\begin{itemize}[leftmargin=*]
    \item \textbf{Hierarchical Structures and Relations}. Existing methods model both user-item interactions and KG in Euclidean space, while both data structures exhibit a highly non-Euclidean latent anatomy. Specifically, the user-item interactions typically follow the power-law distribution (as shown in Figure~\ref{fig:distribution_alibaba}),
    indicating the underlying hierarchical structures~\cite{wsdm20hyperRs,kdd21hyperEmbedding,wsdm22hyperRS}. Meanwhile, hierarchical information is ubiquitous in real-world KGs, since human knowledge is organized hierarchically~\cite{acl20RotH,nips21conE}. Euclidean-based methods are insufficient to capture the intrinsic hierarchical structures of the data, since they suffer from a high distortion when embedding hierarchical data~\cite{nips17poincare,nips18hypernn,nips19hgcn}.
    Moreover, none of them considers KG relations at a finer-grained level of hierarchies. They all conveniently ignore an important fact, \ie hierarchical relations (blue lines in Figure~\ref{fig:kg})
    and non-hierarchical relations (black lines in Figure~\ref{fig:kg})
    are \textbf{not} of equal importance for characterizing items. Taking the outfit recommendation in Figure~\ref{fig:kg} as an example. Hierarchical relation $r_1$ offers complementary information to the cloth $i_1$ in the aspect that $i_1$ is the jeans $e_1$ while non-hierarchical relation $r_5$ means that cloth $i_2$ and cloth $i_3$ can compose an outfit.
    Thus, hierarchical relations can indicate the attributes of items while non-hierarchical relations only reveal the relatedness between entities. Ignoring the hierarchical structures and relations limits model's expressive power.

    \item \textbf{High-order Collaborative Signals of Items.} In existing studies, the item aggregation schemes are mostly KG-oriented, that is, recursively collecting items' knowledge associations from KG without considering the collaborative signals from users~\cite{sigir20ckan,www21kgin}, or blindly mixing the heterogeneous information from neighboring nodes (users or entities) in the Unified Knowledge Graph (UKG)~\cite{kdd19kgat,cikm21kcan}. They fail to explicitly encode the crucial high-order collaborative signals of items 
    , which are latent in user-item interactions and play an important role in learning user preference from behavior aspect.
    Take Figure~\ref{fig:kg} as an example. The path $i_1 \rightarrow u_1 \rightarrow i_3 \rightarrow u_2 \rightarrow i_2$ indicates the long-range connectivity between item $i_1$ and item $i_2$: since users $u_1$ and $u_2$ share the same interest (\ie they both like item $i_3$), the items $i_1$ and $i_2$ which are favored by user $u_1$ and user $u_2$ respectively may be similar to some extent. 
    Therefore, such KG-oriented aggregation schemes are insufficient to explicitly capture the high-order collaborative signals for comprehensive item representations.
\end{itemize}

\begin{figure}[t]
	\centering
	\vspace{-2pt}
	\includegraphics[width=0.175\textwidth]{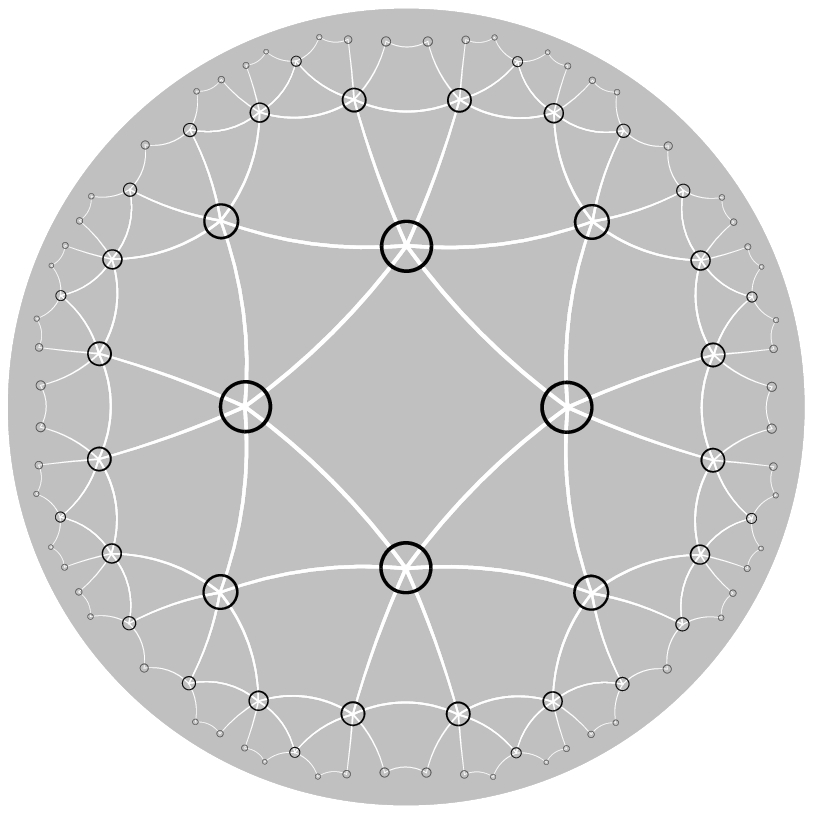}
	\hspace{5.0mm}
	\includegraphics[width=0.175\textwidth]{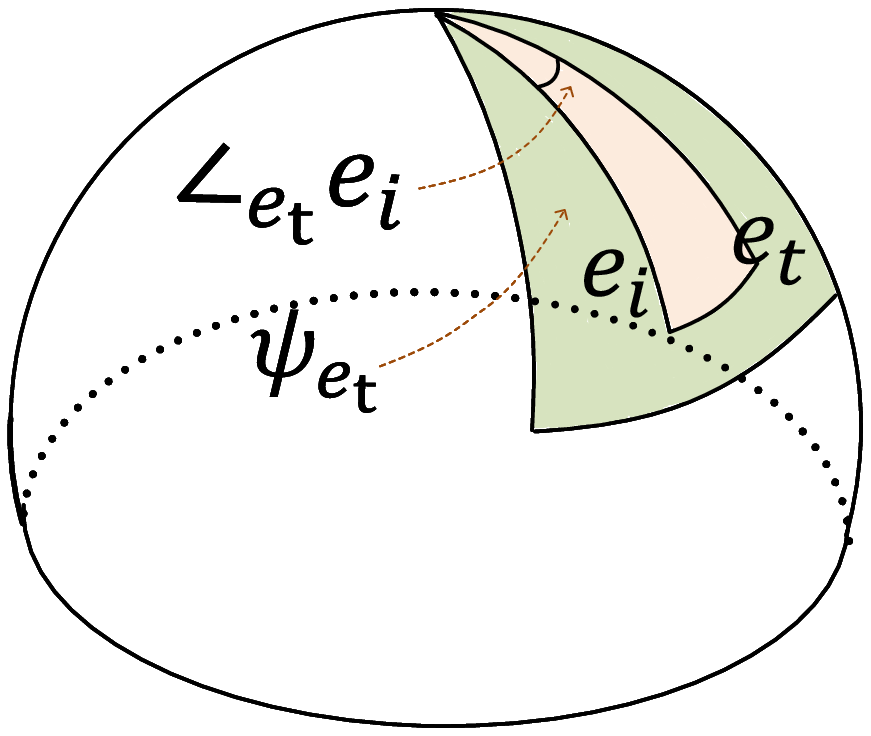}
	\vspace{-2mm}
	\caption{(a) Hyperbolic space expands exponentially. (b) Illustration of Hyperbolic cones.}
	\label{fig:hyper_ball_and_cones}
	\vspace{-13pt}
\end{figure}

To tackle the aforementioned challenges, a new latent space with a smaller embedding distortion is required. Hyperbolic geometry offers an ideal alternative, as it not only enables embeddings with much smaller distortion~\cite{nips18hypernn,nips19hgcn} but also naturally preserves the data hierarchy~\cite{acl20RotH,kdd21hyperEmbedding}. The key property of hyperbolic space is that, unlike Euclidean space, it expands exponentially rather than polynomially. This allows hyperbolic space to have much larger data capacity than Euclidean space, and hence, it has ``sufficient'' room to better preserve the distance between hierarchical data, as shown in the left of Figure~\ref{fig:hyper_ball_and_cones}. Besides, hyperbolic space can be viewed as a continuous version of trees, which makes it perfect to model hierarchical tree-like data due to their analogous structure. Thus, we propose \underline{H}ierarchy-\underline{A}ware \underline{K}nowledge \underline{G}ated Network (HAKG), a new hyperbolic knowledge-aware model with two key components, which can effectively capture and model the aforementioned two important factors that are \textit{not} modeled by any existing methods.

\begin{itemize}[leftmargin=*]
    \item \textbf{Hierarchy-Aware Modeling.} To better capture the underlying hierarchical structures, we map user and item embeddings as well as entity and relation embeddings to hyperbolic space. We also design a new hyperbolic relation-transitive aggregation mechanism to capture the relation dependencies carried by neighbors in hyperbolic space. Moreover, a novel angle constraint is introduced for hierarchical relations, which is able to better preserve items' attributes information in the embedding space and thus improve the representational capacity and expressiveness.

    \item \textbf{Gated Aggregation with Dual Embeddings.}
    Unlike previous KG-oriented aggregation strategies, we view user-item interactions and KG as two different information channels, and develop different aggregation strategies for the two channels. Besides, since items can serve as a natural bridge to connect the two information channels, we use \textit{dual} embedding instead of \textit{single} embedding for items to represent and propagate each channel's information separately, so as to better capture the holistic semantics of items. Moreover, an information gated mechanism is introduced to adaptively fuse the two types of semantics for better identifying user behavior patterns.
    \end{itemize}

To this end, the newly proposed HAKG framework is designed to i) leverage the expressiveness of hyperbolic geometry for better hierarchical modeling, and ii) effectively capture the holistic semantics of items. We conduct extensive experiments on three real-world datasets to evaluate the performance of HAKG and existing methods. Experimental results show that our HAKG significantly outperforms all the start-of-the-art methods such as CKAN~\cite{sigir20ckan}, Hyper-Know~\cite{aaai21hyperKGRec}, and KGIN~\cite{www21kgin}. Furthermore, HAKG is able to 
reveal the underlying hierarchical structures and relations of data in the embedding space for better model expressiveness. 

In summary, our contributions are as follows:

\begin{itemize}[leftmargin=*]
    \item We present knowledge-aware recommendation from a new perspective by taking the hierarchy and high-order items' collaborative signals into consideration.
    \item We embed users and items, as well as entities and relations in hyperbolic space, and design a new hyperbolic aggregation in KG to preserve the relation dependencies among neighbors. Besides, an angle constraint is introduced for hierarchical relations to profile items with the attributes in the embedding space.
    \item We devise the dual item embeddings design to i) better represent and propagate collaborative signals and knowledge associations simultaneously, and ii) more effectively leverage the information gate mechanism to control discriminative signals towards the users' preference from both behavior and attribute aspects.
    \item We conduct extensive experiments on three public benchmark datasets to demonstrate the superiority of HAKG.
\end{itemize}

\section{Problem Formulation}
\label{sec:problem_formulation}

We first introduce the data structures related to our studied problem, and then formulate our task.

\vspace{3pt}
\noindent
\textbf{User-Item Bipartite Graph.}
In this paper, we focus on learning the user preference from implicit feedback~\cite{uai09BPRloss}. To be more specific, the behavior data (\eg click and review) involves a set of users $\Set{U} = \{u\}$ and a set of items $\Set{I} = \{i\}$. We model user-item interactions as a bipartite graph $\Set{G}_b = \{(u,i) | u\in\Set{U},i\in\Set{I}\}$ where each $(u,i)$ pair indicates that user $u$ has interacted with item $i$.

\vspace{3pt}
\noindent\textbf{Knowledge Graph (KG).}
KGs are collections of real-world facts, such as item attributes, concepts, or external commonsense. Let $\Set{T}$ be the triplet set, $\Set{E}$ be a set of entities, and $\Set{R}$ be the relation set, which involves relations in both canonical and inverse directions (\eg \textit{compose} and \textit{composed-of}). Let KG be a heterogeneous graph $\Set{G}_k= \{(h,r,t) | h,t\in \Set{E}, r\in \Set{R}\}$,  where each triplet $(h,r,t) \in \Set{T}$ means that there is a relation $r$ between head entity $h$ and tail entity $t$.
For example, a triplet (\textit{jeans}, \textit{brand}, \textit{Levi's}) indicates that the \textit{brand} of the \textit{jeans} is \textit{Levi's}.
As we assume all the items appear in KG as entities (i.e., $\Set{I} \subset \Set{E}$), a common assumption made by all existing knowledge-aware recommendation systems~\cite{www19kgcn,kdd19kgat,www21kgin}, we can link items in user-item graph with entities in KG to offer auxiliary semantics to interactions.
%

\vspace{3pt}
\noindent\textbf{Task Description.}
Given a user-item graph $\Set{G}_b$ and a KG $\Set{G}_k$, our task of knowledge-aware recommendation is to predict how likely that a user would adopt an item that she has never engaged with.

\section{Methodology}
\label{sec:methodology}

\begin{figure*}[t]
    \vspace{-15pt}
	\centering
	\vspace{-15pt}
	\includegraphics[width=0.90\textwidth]{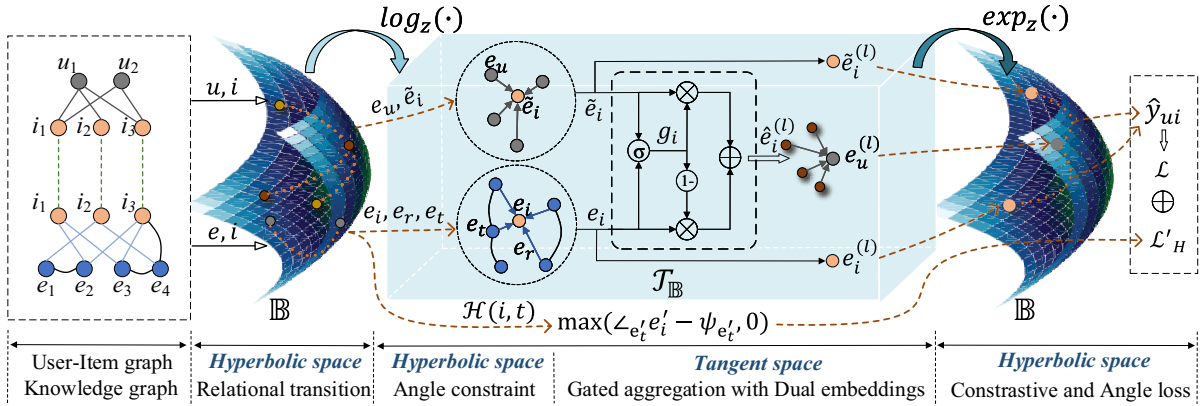}
	\vspace{-10pt}
	\caption{Illustration of the proposed HAKG framework.}
	\label{fig:HAKG-overview}
	\vspace{-10pt}
\end{figure*}

In this section, we first introduce the basic concepts about hyperbolic space, and then detail two key components of HAKG, \ie \textit{hierarchy-aware modeling} and \textit{gated aggregation with dual embeddings}. The former embeds users and items as well as entities and relations in hyperbolic space. It adopts a novel hyperbolic relation-aware aggregation over KG, and also introduces an angle constraint for learning attribute semantics of item in the embedding space. The latter leverages dual item embeddings to separately encode the high-order collaborative signals from user-item graph and the knowledge associations from KG, and distills useful semantics via the gate mechanism for learning high-quality user representations.

\subsection{Preliminaries}
\label{sec:Preliminaries}
Hyperbolic geometry~\cite{Cannon97hyperbolicgeometry} is a non-Euclidean geometry with constant negative curvature. Compared with the Euclidean spaces, the amount of space covered by the hyperbolic geometry increases exponentially rather than polynomially \wrt the radius. This property allows us to effectively capture underlying hierarchical structures of user-item interactions and KG in hyperbolic space. In this work, we use the Poincar\'e ball model with constant curvature $-c=-1$ for learning hyperbolic embeddings due to its feasibility for gradient optimization~\cite{nips17poincare,kdd21hyperEmbedding}. In the following, we first introduce the necessary mathematical basis for the Poincar\'e ball model.

\noindent
\textbf{Poincar\'e Ball \& Tangent Space.} The definition domain of the Poincar\'e ball model is:
{\setlength{\abovedisplayskip}{3pt}
\setlength{\belowdisplayskip}{3pt}
\begin{gather}\label{equ:poincare-ball}
\Space{B}=\{(x_{1}, \ldots, x_{n}): x_{1}^{2}+\cdots+x_{n}^{2}<\frac{1}{c}\}
\end{gather}
}in $\Space{R}^n$. The tangent space $\Set{T}_\Mat{z}\Space{B}$ at point $z$ on $\Space{B}$ is a $n$-dimensional Euclidean space that best approximates $\Space{B}$ around $z$, which is useful to perform aggregation operations in hyperbolic space~\cite{nips19hgcn}.

\noindent
\textbf{Exponential Map \& Logarithmic Map.} Hyperbolic space and
tangent space can be bridged by exponential and logarithmic mappings. Specifically, the exponential map can map the tangent space $\Set{T}_\Mat{z}\Space{B}$ to the hyperbolic space $\Space{B}$, and the logarithmic map maps  $\Space{B}$ to $\Set{T}_\Mat{z}\Space{B}$ conversely. In particular, the closed-form expressions for the two maps at point $z$ are defined below.
{\setlength{\abovedisplayskip}{3pt}
\setlength{\belowdisplayskip}{3pt}
\begin{gather}\label{equ:hyperbolic-map}
\exp_{\Mat{z}}(\Mat{x})=\Mat{z} \oplus \tanh (\frac{\|\Mat{x}\|}{1-\|\Mat{z}\|^2}) \frac{\Mat{x}}{\|\Mat{x}\|}\\ \label{equ:log-map}
\log_{\Mat{z}}(\Mat{y})=(1-\|\Mat{z}\|^2) \cdot \tanh ^{-1}(\|-\Mat{z} \oplus \Mat{y}\|) \frac{-\Mat{z} \oplus \Mat{y}}{\|-\Mat{z} \oplus \Mat{y}\|}
\end{gather}
where $\oplus$ represents M\"obius addition~\cite{nips18hypernn}:
\begin{gather}\label{equ:mobius-addition}
\Mat{x} \oplus \Mat{y}=\frac{(1+2\langle \Mat{x}, \Mat{y}\rangle+\|\Mat{y}\|^{2}) \Mat{x}+(1-\|\Mat{x}\|^{2}) \Mat{y}}{1+2\langle \Mat{x},  \Mat{y}\rangle+\|\Mat{x}\|^{2}\|\Mat{y}\|^{2}}
\end{gather}}

\subsection{Hierarchy-Aware Modeling}
\label{sec:hierarchy_modeling}

Unlike the previous propagation-based studies~\cite{kdd19kgat,kdd19kgnn-ls,cikm21kcan,www21kgin} that simply embed user-item graph and KG in Euclidean space, we aim to capture the non-Euclidean latent anatomy of data. Taking the Alibaba-iFishion dataset shown in Figure~\ref{fig:distribution_alibaba}
as an example, the degree distribution of user-item interactions reveals an underlying hierarchical tree-like structure with power-law distribution.
Meanwhile, KG also exhibits hierarchical patterns~\cite{acl20RotH,nips21conE}. However, the hierarchical property of data cannot be properly captured in Euclidean space because of the high-distortion~\cite{nips17poincare,nips18hypernn}. Thus, we embed users and items as well as entities and relations in hyperbolic space, where the hierarchical structures can be naturally preserved. We also design a new aggregation scheme in KG since previous Euclidean-based methods are not feasible in hyperbolic space.

Meanwhile, we argue that previous methods are unable to characterize items properly because they do not differentiate hierarchical relations from the rest, and \textit{only} model KG relations at a coarse granularity.
Specifically, hierarchical relations in KG often carry entities' attributes information (\eg \textit{brand} and \textit{fabric}),
while non-hierarchical relations can only indicate the relatedness between entities (\eg \textit{match} and \textit{similar to}).
This motivates us to introduce an additional constraint for hierarchical triplets in the embedding space to improve the representational capacity and expressiveness.

\subsubsection{\textbf{Hyperbolic Relation-Transitive Aggregation.}} We first consider the aggregation scheme of KG in hyperbolic space. As an item $i$ can be involved in multiple KG triplets, 
its neighborhood reflects the relational similarity between $i$ and its connected entities to a certain degree.
Formally, given a target item $i$ in KG, aggregating local information from $i$'s neighbors in KG can reveal related knowledge associations for $i$. Thus, we use $\Set{N}_{i}=\{(r,t)|(i,r,t)\in\Set{G}_k\}$ to represent the neighborhood entities and the first-order relations of item $i$ in KG, and propose to integrate the relational context from neighborhood entities to generate the \emph{knowledge representation} of item $i$:
{\setlength{\abovedisplayskip}{3pt}
\setlength{\belowdisplayskip}{3pt}
\begin{gather}\label{equ:1-hop-kg-aggregation}
    \Mat{e}^{(1)}_{i}=f_{\Space{B}}\Big(\{(\Mat{e}^{(0)}_{i},\Mat{e}_{r},\Mat{e}^{(0)}_{t})|(r,t)\in\Set{N}_{i}\}\Big)
\end{gather}}where $\Mat{e}^{(1)}_{i}\in\Space{B}^{d}$ is the hyperbolic knowledge representation that collects the contextual information of item $i$
from the first-order connectivity, and $f_{\Space{B}}(\cdot)$ is the hyperbolic aggregation function to extract and integrate information related to item $i$
from each connection $(i,r,e)$ in hyperbolic space. Previous studies~\cite{www21kgin} have shown that node-based aggregation cannot differentiate relational paths, which fails to preserve the relation dependencies carried by paths. Consequently, it is necessary to integrate relational paths into representations. Besides, Euclidean propagation-based methods mainly use mean aggregation to average the neighbor information. An analog of mean aggregation in hyperbolic space is the Fr\'echet mean~\cite{frechetmean1948}, which, however, has no closed form solution. Instead, we model relational context in the mean aggregator as:
\begin{gather}\label{equ:kg-aggregation}
    \Mat{e}^{(1)}_{i}=\exp_{\Mat{0}}\Big( \frac{1}{|\Set{N}_{i}|}\sum\nolimits_{(r,t)\in\Set{N}_{i}}\log_{\Mat{e}_i^{(0)}}(\Mat{e}_{t}^{(0)}\oplus\Mat{e}_{r})\Big)
\end{gather}
where $\oplus$ is the M\"obius addition operation, $\log(\cdot)$ and $\exp(\cdot)$ are the aforementioned exponential map and logarithmic map respectively, and $\Mat{e}^{(0)}_{t}$ is the ID embedding of entity $t$ in hyperbolic space. For each triplet $(i,r,t)$ in KG, we devise a relational transition $\Mat{e}_{t}^{(0)}\oplus\Mat{e}_{r}$ by modeling the relation $r$ as the vector translation of $\Mat{e}^{(0)}_{t}$. To avoid the complex mean operation in hyperbolic space, we first map each relational context
to the tangent space, as this is where the Euclidean approximation performs the best (\cf Figure~\ref{fig:HAKG-overview}); we then leverage the exponential map to map it back to hyperbolic space to obtain the representation $\Mat{e}^{(1)}_{i}$. As a result, the aggregator is able to integrate relational message into target representations in hyperbolic space, and avoid intricate hyperbolic mean pooling via map operations. Analogously, we can obtain the representation $\Mat{e}^{(1)}_{e}$ of each KG entity $e\in\Set{E}$.

We further stack more aggregation layers to explore the high-order knowledge associations for items. Technically, we recursively formulate the knowledge  representations of item $i$ after $l$ layers as:
\begin{gather}\label{equ:kg-final-aggregation}
    \Mat{e}^{(l)}_{i}=\exp_{\Mat{0}}\Big( \frac{1}{|\Set{N}_{i}|}\sum\nolimits_{(r,t)\in\Set{N}_{i}}\log_{\Mat{e}_i^{(l-1)}}(\Mat{e}_{t}^{(l-1)}\oplus\Mat{e}_{r})\Big)
\end{gather}

\subsubsection{\textbf{Angle Constraint of Hierarchical triplets.}}
\label{sec:angle-constraint}
Next, we focus on modeling the attributes information of items with hierarchical relations. To ease our discussion, we assume that all the relations in KG are clustered into two types (\ie hierarchical and non-hierarchical), and the types of relations are given \textit{a priori}. In fact, the types of relations are explicitly available in most KGs~\cite{acl20RotH,nips21conE}. When such information is not available, there are two alternative options. One is to infer the ``hierarchicalness'' of a relation by widely used Krackhardt scores criteria~\cite{krackhardt2014graph}; the other is to simply assume that item-connected entities are acted as the attributes of items, and thus, these relations are hierarchical. Our experiments to be reported in Section~\ref{sec:exp_hierarchiy} show that the difference between two options is negligible.

As mentioned earlier, hierarchical relations can reveal the attributes of items. However, above aggregation scheme cannot preserve this semantic in the embedding space since it simply aggregates all neighboring information into representations. To explicitly model items' characteristic information, we consider the hyperbolic entailment cone~\cite{icml18cone}, which is a family of convex cones that can model the hierarchies with nested angular cones. Technically, we use $\Set{C}_{\Mat{x}}$ to denote the cone at apex $\Mat{x}$, and define the cone width function  (\ie half aperture\footnote{\url{https://en.wikipedia.org/wiki/Cone}} of cone $\Set{C}_{\Mat{x}}$) as $\psi_{\Mat{x}} = \arcsin (K \frac{1-\|\Mat{x}\|^{2}}{\|\Mat{x}\|})$, where $K\in\Space{R}$ is a hyperparameter (we follow~\cite{nips21conE} and set $K = 0.1$). Also, for $\Mat{x},\Mat{y}\in\Space{B}^d$, we define the angle of $\Mat{y}$ at $\Mat{x}$ to be the angle between the half-lines $\overrightarrow{\Mat{ox}}$ and $\overrightarrow{\Mat{xy}}$, and denote it as $\angle_\Mat{x} \Mat{y}$:
\begin{gather}\label{equ:cone-angle}
\angle_\Mat{x} \Mat{y}=\arccos (\frac{\langle\Mat{x}, \Mat{y}\rangle(1+\|\Mat{x}\|^{2})-\|\Mat{x}\|^{2}(1+\|\Mat{y}\|^{2})}{\|\Mat{x}\|\|\Mat{x}-\Mat{y}\| \sqrt{1+\|\Mat{x}\|^{2}\|\Mat{y}\|^{2}-2\langle\Mat{x}, \Mat{y}\rangle}})
\end{gather}
The transitivity property of angular cones~\cite{icml18cone} guarantees that if $\Mat{y}$ is in $\Set{C}_{\Mat{x}}$,  $\Set{C}_{\Mat{y}}$ is also in $\Set{C}_{\Mat{x}}$. In other words, hyperbolic entailment cone forms a nested structure in embedding space, and the width (\aka angle) of the cone can naturally indicate the \textit{``attribute''} semantics of embeddings, as shown in the right of Figure~\ref{fig:hyper_ball_and_cones}. To leverage the expressive power of hyperbolic entailment cone geometry, we first use $\Set{H}=\{(i,t)|(i,r,t)\in\Set{G}_k \land r \text{ is hierarchical}\}$ to represent the set of hierarchical entities in terms of hierarchical relations $r$ in KG, and use Eq.~(\ref{equ:angle-loss}) to define the angle constraint in embedding space:
\begin{gather}\label{equ:angle-loss}
    \Lapl_{\text{H}} = \sum\nolimits_{(i,t)\in\Set{H}}\max(\angle_{\Mat{e}_{t}} \Mat{e}_i - \psi_{\Mat{e}_t}, 0)
\end{gather}
where $\Mat{e}_{t}$ and $\Mat{e}_i$ are the final knowledge representations of entity $i$ and entity $t$ respectively, defined in Eq.~(\ref{equ:final-representations}). The hinge loss on angle constraint suggests the angle of $\angle_{\Mat{e}_{t}}\Mat{e}_i$ to be smaller than the width of entailment cone $\Set{C}_{\Mat{t}}$, meaning that the embedding of item $\Mat{e}_i$ should be ``contained'' inside of the embedding of entity $\Mat{e}_t$ in the entailment cone geometry.

While such a constraint can ensure the attributes semantic with geometric angles, it also limits the representational capacity. For example, let item $i$ be a jeans with multiple hierarchical relations (\eg \textit{brand} and \textit{fabric}). As each of $i$'s connected entity defines an entailment cone in hyperbolic space, the embedding of $i$ can only be in their intersection area. Since items often have dozens of relations in KG, this intersection area could be extremely small, leading to possible collapse in the embedding space. Thus, we relax it by randomly assigning a corresponding subspace for the embedding of item $i$
that satisfies the following constraint:
\begin{gather}\label{equ:angle-mask-loss}
    \Lapl_{\text{H}}^{\prime} = \sum\nolimits_{(i,t)\in\Set{H}} \max(\angle_{\Mat{e}_{t}^\prime} \Mat{e}_i^\prime - \psi_{\Mat{e}_t^\prime}, 0)
\end{gather}
where $\Mat{e}_{i}^\prime$ and $\Mat{e}_{t}^\prime$ are the subspace embeddings of $\Mat{e}_{i}$ and $\Mat{e}_{t}$ respectively by randomly masking some dimensions. In our current implementation, each dimension has a probability of 0.5 to be masked off, while this probability can be easily adjusted to cater for the application needs.
By doing so, we only enforce the angle constraint in a subset of $d$ hyperbolic planes, leaving room for expressing relatedness and other knowledge semantics of items.

\subsection{Gated Aggregation with Dual embeddings}
\label{sec:gate_propagation}

The second key component of HAKG framework is designed mainly to capture the high-order collaborative signals of items.
Existing methods utilize KG-oriented aggregation strategies
to collect the information related to an item from its neighboring nodes. Unfortunately, this approach either mixes the collaborative signals with knowledge associations such that they become indistinguishable~\cite{kdd19kgat,wsdm22hyperRS}, or completely neglects the collaborative signals of items~\cite{sigir20ckan,www21kgin} that results in suboptimal item representations. Thus, we propose to separately aggregate the two types of information with dual item embeddings to solve the limitation: (1) Knowledge item embeddings, which are used to characterize items with knowledge information from KG, as demonstrated in Section~\ref{sec:hierarchy_modeling}. (2) Collaborative item embeddings, which encode items' high-order collaborative signals such as co-occurrence relationship in user-item graph. Neighbored with the holistic representations of items, we further develop an information gate mechanism to adaptively distill useful information for users' aggregation. We illustrate our approach in Figure~\ref{fig:HAKG-overview}.



\subsubsection{\textbf{Collaborative Aggregation for Items.}} To explicitly encode the collaborative signals of items, we initialize new item representations $\tilde{\Mat{e}}^{(0)}$ in hyperbolic space to items' collaborative information, which are latent in user-item interactions. We also use the neighborhood aggregation scheme to integrate an item's multi-hop neighbors into its representation for capturing the high-order connectivities. Given an item $i$ in the user-item graph, we use $\tilde{\Set{N}_{i}}=\{u|(u,i)\in\Set{G}_b\}$ to represent all the users who have interacted with the item. To generate the $l$-th layer representation of item $i$, we recursively integrate the collaborative information from item $i$'s neighbor users:
{\setlength{\abovedisplayskip}{3pt}
\setlength{\belowdisplayskip}{3pt}
\begin{gather}\label{equ:item-aggregation}
    \tilde{\Mat{e}}^{(l)}_{i}=\exp_{\Mat{0}}\Big( \frac{1}{|\tilde{\Set{N}_{i}}|}\sum\nolimits_{(u)\in\tilde{\Set{N}_{i}}}\log_{\Mat{0}}(\Mat{e}_u^{(l-1)})\Big)
\end{gather}}where $\tilde{\Mat{e}}^{(l)}_{i}$ is the collaborative embedding of item $i$ after $l$ layers, $\log(\cdot)$ and $\exp(\cdot)$ are the exponential map and logarithmic map that are used to map between Euclidean space and hyperbolic space, and $\Mat{e}_u^{(l-1)}$ is the hyperbolic embedding of user $u$ after $l-1$ layers in hyperbolic space, which we will detail below.

\subsubsection{\textbf{Information Gated Aggregation for Users.}} To distill useful information from neighbor items, we design a gated module which is able to generate discriminative signals for users' aggregation. Specifically, inspired by the design of GRU~\cite{14gru} that learns gating signals to control the update of hidden states, we propose to learn a fusion gate to adaptively control the combination of two different types of semantic item representations. For an item $i$, given its knowledge representation ${\Mat{e}}^{(l)}_{i}$ and collaborative representation $\tilde{\Mat{e}}^{(l)}_{i}$ in the $l$-th layer, we balance the contributions of the two different types of information with a learnable gating fusion unit:
{\setlength{\abovedisplayskip}{3pt}
\setlength{\belowdisplayskip}{3pt}
\begin{gather}\label{equ:gate}
    \Mat{g}^{(l)}_{i}=\sigma(W_1 \log_{\Mat{0}}({\Mat{e}}^{(l)}_{i}) + W_2 \log_{\Mat{0}}(\tilde{\Mat{e}}^{(l)}_{i})) \\
   \hat{\Mat{e}}^{(l)}_{i} = \exp_{\Mat{0}}\Big(\Mat{g}^{(l)}_{i}\cdot{\Mat{e}}^{(l)}_{i} + (1-\Mat{g}^{(l)}_{i})\cdot\tilde{\Mat{e}}^{(l)}_{i}\Big)
\end{gather}}where $W_1, W_2 \in\Space{R}^{d\times d}$ are learnable transformation parameters, and $\sigma(\cdot)$ is the Sigmoid function. Notation $\Mat{g}^{(l)}_{i}\in\Space{R}^d$ denotes the learned gate signals to balance the contributions of collaborative signals and knowledge association of item $i$, as shown in Figure~\ref{fig:HAKG-overview}. A high value of $\Mat{g}^{(l)}_{i}$ indicates that the users interact with item $i$ mainly because of the attributes of item $i$ (\eg $i$ is the jeans), rather than the behavior similarities among items (\eg the jeans $i$ liked by user $u$ is collaboratively relevant to other bottoms that user $u$ likes).
Let $\Set{N}_{u}=\{i|(u,i)\in\Set{G}_b\}$ denote the set of items that user $u$ has interacted with. Then, we can formulate the representation of user $u$ with distilled information:
\begin{gather}\label{equ:user-aggregation}
    \Mat{e}^{(l)}_{u}=\exp_{\Mat{0}}\Big( \frac{1}{|\Set{N}_{u}|}\sum\nolimits_{(i)\in\Set{N}_{u}}\log_{\Mat{0}}(\hat{\Mat{e}}_i^{(l-1)})\Big)
\end{gather}

\subsection{Model Prediction}

After $L$ layers, we obtain the representations of item $i$ and user $u$ at different layers, and then sum them up as the final representation:
\begin{gather}\label{equ:final-representations}
    \Mat{e}_{i}=\sum\nolimits_{l=0}^L \Mat{e}_{i}^{(l)}, \qquad
    \tilde{\Mat{e}}_{i}=\sum\nolimits_{l=0}^L \tilde{\Mat{e}}_{i}^{(l)}, \qquad
    \Mat{e}_{u}=\sum\nolimits_{l=0}^L \Mat{e}_{u}^{(l)}
\end{gather}
By doing so, the complementary information of collaboration and knowledge is separately encoded in the final representations. 
Different from previous work, we employ the consine similarity to separately predict how likely the user $u_i$ would adopt an item $i$ from item' behavior and attributes aspects, and then adopt the sum of the two predictions as the final prediction score $\hat{y}_{ui}$:
\begin{gather}\label{equ:prediction}
    \hat{y}_{ui} = \cos(e_u, \tilde{e}_i) +  \cos(e_u, e_i)
\end{gather}

\subsection{Model Optimization}
\label{sec:model_optimization}
We opt for the contrastive loss~\cite{cikm21simpleX} to optimize HAKG. Compared with widely used BPR loss~\cite{uai09BPRloss}, it alleviates the convergence problem by introducing more negative samples and penalizing uninformative ones. Specifically, for each interaction $(u,i)$ captured by user-item graph (\ie positive pair), we randomly sample |$\Set{M}_u$| unobserved items to form the negative pairs together with user $u$, denoted as $\Set{M}_u$, and maximize the similarity between the positive pair and meanwhile minimize the similarity of negative pairs with a margin $m$: 
\begin{gather}\label{equ:contrastive-loss}
    \Lapl = \sum\nolimits_{(u,i)\in\Set{D}}\big[2-\hat{y}_{ui} + \frac{1}{|\Set{M}_u|}\sum\nolimits_{j\in\Set{M}_u}(\hat{y}_{uj}-m)_{+}\big]
\end{gather}
where $\Set{D}$
is the interaction data, and $(\cdot)_{+}$ is the ramp function $\max(0,\cdot)$. By combing the prediction loss and angle loss, we minimize the following objective function to learn the model parameters:
\begin{gather}\label{equ:final-loss}
    \Lapl =  \Lapl + \lambda \Lapl_{\text{H}}^\prime
\end{gather}
where $\lambda$ is a hyperparameter to control the weight of angle loss defined in Eq.~(\ref{equ:angle-loss}).


\subsection{Model Analysis}
\label{sec:model_analysis}

\subsubsection{\textbf{Model Size.}} Previous studies~\cite{www21kgin} have confirmed that discarding the nonlinear feature transformations not only yields better performance but also reduces redundant parameters. Hence, in the aggregation schemes of HAKG, we discard the nonlinear activation function and the learnable transformation metrics. The model parameters of HAKG consist of (1) ID embedding of users, items, KG entities and relations $\{\Mat{e}_u^{(0)},\tilde{\Mat{e}}_i^{(0)},\Mat{e}_e^{(0)},\Mat{e}_r|u\in\Set{U},i\in\Set{I},e\in\Set{E},r\in\Set{R}\}$; and (2) transformation parameters $W_1, W_2$ for information gated aggregation used in Eq.~(\ref{equ:gate}).

\subsubsection{\textbf{Time Complexity.}} The time cost of HAKG mainly comes from the aggregation schemes. In the aggregations over KG, the computational complexity of updating knowledge item embeddings in hyperbolic space is $O(|\Set{G}_k|dL)$, where $|\Set{G}_k|$, $d$, and $L$ denote the number of KG triplets, the embeddings size, and the number of layers. In the aggregation over user-item graph, the computational complexity of collaborative item embeddings and user embeddings is $O(|\Set{G}_b|dL)$, where $|\Set{G}_b|$ is the number of interactions. For the hierarchy-aware modeling, the cost of angle constraint is $O(|\Set{I}|)$, where $|\Set{I}|$ is the number of items. Besides, the cost of mappings between Euclidean space and hyperbolic space is $O(|\Set{G}_b| + |\Set{G}_k|)$. Thus, the time complexity of the whole training epoch is $O(|\Set{G}_k|dL + |\Set{G}_b|dL)$. Under the same experimental settings (\ie same embeddings size and propagation layer), HAKG has comparable complexity to KGAT, CKAN, and KGIN, three representative propagation-based methods.

\section{Experiments}
\label{sec:expperiments}
We present empirical results to demonstrate the effectiveness of our proposed HAKG framework. The experiments are designed to answer the following three research questions:
\begin{itemize}[leftmargin=*]
    \item \textbf{RQ1:} How does HAKG perform, compared with the state-of-the-art knowledge-aware recommendation models?
    \item \textbf{RQ2:} How do different components of HAKG (\ie hierarchical modeling, gated propagation, dual embeddings, and the depth of propagation layers) affect the performance of HAKG?
    \item  \textbf{RQ3:} Can HAKG provide meaningful insights on the hierarchical modelling both from structures and relations?
\end{itemize}

\subsection{Experimental Settings}
\label{sec:experimental_settings}

\subsubsection{\textbf{Dataset Description.}}
\label{sec:dataset_description}
We utilize three benchmark datasets to evaluate the performance of HAKG: Alibaba-iFashion, Yelp2018, and Last-FM. These three datasets are widely adopted in the state-of-the-art methods~\cite{kdd19kgat, www21kgin, aaai21hyperKGRec,wsdm22hyperRS}, and vary in terms of domain, size, and sparsity. Following previous work~\cite{kdd19kgat}, we collect the two-hop neighbor entities of items in KG to construct the item knowledge graph for each dataset. Besides, for the hierarchical types of relations in KG, without loss of generality, we simply treat the item-connected relations are hierarchical, since they are able to reveal the attributes of items. Moreover, in order to ensure the data quality, we adopt the 10-core setting~\cite{kdd19kgat}, \ie retaining users and items with at least ten interactions and filtering out KG entities involved in less than ten triplets. We use the same data partition with previous studies~\cite{kdd19kgat,www21kgin} for comparison (\ie the proportions of training, validation, and testing set are 80\%, 10\%, and 10\% for all datasets). Table~\ref{tab:dataset} summarizes the overall statistics of the three datasets used in our experiments.

\subsubsection{\textbf{Evaluation Metrics.}} We adopt the all-ranking strategy~\cite{kdd20sampled,www21kgin} to evaluate the performance.
Specifically, we treat all the items that user has not adopted before as negative, and treat the relevant items in the testing set as positive.
To evaluate the effectiveness of top-$K$ recommendation, we adopt two widely-used evaluation protocols~\cite{www17ncf,kdd20sampled} recall@$K$ and ndcg@$K$, where $K = 20$ by default. We report the average metrics for all the users in the testing set.

\begin{table}[t]
    \caption{Statistics of the datasets used in our experiments.}
    \vspace{-10pt}
    \label{tab:dataset}
    \resizebox{0.47\textwidth}{!}{
    \begin{tabular}{c|l|r|r|r}
    \hline
    \multicolumn{1}{l|}{} &  & \multicolumn{1}{c|}{Alibaba-iFashion} & \multicolumn{1}{c|}{Yelp2018} & \multicolumn{1}{c}{Last-FM}  \\ \hline\hline
    \multirow{3}{*}{\begin{tabular}[c]{@{}c@{}}User-Item\\ Interaction\end{tabular}} & \#Users & 114,737 & 45,919 & 23,566 \\
     & \#Items & 30,040 & 45,538 & 48,123  \\
     & \#Interactions & 1,781,093 & 1,185,068 & 3,034,796  \\ \hline\hline
    \multirow{3}{*}{\begin{tabular}[c]{@{}c@{}}Knowledge\\ Graph\end{tabular}} & \#Entities & 59,156 & 90,961 & 58,266  \\
     & \#Relations & 51 & 42 & 9  \\
     & \#Triplets & 279,155 & 1,853,704 & 464,567  \\ \hline
    \end{tabular}}
    \vspace{-10pt}
\end{table}

\subsubsection{\textbf{Baselines.}} In order to demonstrate the effectiveness of HAKG, we compare it with the state-of-the-art methods, including KG-free method (MF), embedding-based methods (CKE and UGRec), propagation-based methods (KGNN-LS, KGAT, CKAN, and KGIN), and hyperbolic-based methods (Hyper-Know and LKGR):
\begin{itemize}[leftmargin=*]
    \item \textbf{MF}~\cite{uai09BPRloss} is a benchmark factorization model, which only considers the user-item interactions, leaving KG untouched. Here, we use ID embeddings of users and items to perform the prediction.
    \item \textbf{CKE}~\cite{kdd16cke} is a representative embedding method, which utilizes TransR~\cite{aaai15transR} to learn item structural representations from KG, and feeds learned embeddings to MF in an integrated framework.
    \item \textbf{UGRec}~\cite{sigir21UGRec} is a state-of-the-art embedding method that models directed and undirected relations from KG and co-occurrence behavior data. While such undirected relations are inaccessible for other methods and unavailable for the three datasets, we add the connectivities between items which are co-interacted by the same user, and treat them as the co-occurrence relationships.
    \item \textbf{KGNN-LS}~\cite{kdd19kgnn-ls} is a propagation-based model, which transforms KG into user-specific graphs, and then considers user preference on KG relations and label smoothness in the information aggregation phase, so as to generate personalized item representations.
    \item \textbf{KGAT}~\cite{kdd19kgat} is a propagation-based recommender model. It applies a unified relation-aware attentive aggregation mechanism in UKG to generate user and item representations.
    \item \textbf{CKAN}~\cite{sigir20ckan} is based on KGNN-LS, which utilizes different aggregation schemes on the user-item graph and KG respectively, to encode knowledge association and collaborative signals.
    \item \textbf{KGIN}~\cite{www21kgin} is a state-of-the-art propagation-based method, which models user interaction behaviors with latent intents, and proposes a relation-aware information aggregation scheme to capture long-range connectivity in KG.
    \item \textbf{Hyper-Know}~\cite{aaai21hyperKGRec} is a state-of-the-art hyperbolic method that embeds KG in hyperbolic space. An adaptive regularization mechanism is also proposed to adaptively regularize items and their neighboring representations.
    \item \textbf{LKGR}~\cite{wsdm22hyperRS} is a state-of-the-art hyperbolic GNN method with Lorentz model, which employs different information propagation strategies in the hyperbolic space to encode heterogeneous information from historical interaction and KG.
\end{itemize}

\subsubsection{\textbf{Parameter Settings.}}
\label{exp:parameter-setting}
We implement HAKG in Pytorch, and have released our implementations\footnote{\url{https://github.com/Scottdyt/HAKG}} (codes, parameter settings, and training logs) to facilitate reproducibility. For a fair comparison, we fix the size of ID embeddings as 64, and optimize all models with Adam~\cite{iclr2015adam} optimizer, where the batch size is fixed at 4096. The Xavier initializer~\cite{aistats10Xavier} is used to initialize the model parameters. We apply a grid search for hyperparameters: the learning rate is tuned amongst $\{10^{-4},10^{-3},10^{-2}\}$, the weight of angle loss $\lambda$ is searched in $\{10^{-5},10^{-4},\cdots,10^{-2},10^{-1}\}$, and tune the GNN layers $L$ in $\{1,2,3\}$ for propagation-based methods. For the number of negative samples $|\Set{M}_u|$ per user and the margin $m$ of contrastive loss, we set to $\{200,400,400\}$ and $\{0.6,0.8,0.7\}$ for Alibaba-iFashion, Yelp2018, and Last-FM datasets, respectively. Besides, since optimization in hyperbolic space is practically challenging, we instead define all parameters in the tangent space at the origin, optimize embeddings using standard Euclidean techniques, and use the exponential map to recover the hyperbolic parameters~\cite{acl20RotH}. The parameters for all baseline methods are carefully tuned to achieve optimal performance. Specifically, for KGAT, we set the depth to three with the hidden size $\{64,32,16\}$, and use the pre-trained ID embeddings of MF as initialization; for UGRec, we set the margin value for hinge loss to 1.5; for CKAN, KGNN-LS, and LKGR, we set the size of neighborhood to 16; for KGIN, we fix the number of intents to 4; and for Hyper-Know, we set the curvature to -1. Moreover, early stopping strategy is performed for all methods, \ie premature stopping if recall@20 on the test set does not increase for 10 successive epochs.

\subsection{Performance Comparison (RQ1)}
\label{sec:performace}

We begin with the performance comparison \wrt recall@20 and ndcg@20. The experimental results are reported in Table~\ref{tab:overall-performance}, where \%Imp. denotes the relative improvement of the best performing method (starred) over the strongest baselines (underlined). We have the following observations:

\begin{itemize}[leftmargin=*]
    \item HAKG consistently yields the best performance on all the datasets. In particular, it achieves significant improvement even over the strongest baselines \wrt ndcg@20 by 15.43\%, 8.21\%, and 9.79\% in Alibaba-iFashion, Yelp2018, and Last-FM, respectively. We attribute these improvement to the hierarchical modeling and holistic information propagation of HAKG: (1) By embedding user-item interactions and KG in hyperbolic space and introducing auxiliary angle constraint for hierarchical relations, HAKG is able to capture the underlying hierarchical structure of data, and better characterize items in the embedding space. In contrast, all baselines ignore the importance of hierarchical relations, and simply use KG-oriented aggregation schemes to capture KG information. (2) Benefited from our gated aggregation with dual item embeddings, HAKG can explicitly preserve the holistic semantics of items to better encode user behavior patterns, while other propagation-based baselines (\eg CKAN, KGIN, and LKGR) fail to capture the high-order collaborative signals of items.
    \item Jointly analyzing the performance of HAKG across the three datasets, we find that the improvement on Alibaba-iFashion dataset is more significant than that on other datasets. One possible reason is that the size of KG on Alibaba-iFashion is much smaller than that on Yelp2018 and Last-FM, and most of the relations are hierarchical, thus it is more important to i) preserve the valuable attribute information of items, and ii) capture the crucial collaborative signals of items for learning user preference. This confirms that HAKG can better leverage the user-item interactions and KG for comprehensive item representations.
    \item The side information of KG is important for recommendations. Compared with vanilla MF, CKE outperforms MF by simply incorporating KG embeddings into MF. The results are consistent with prior studies~\cite{www21kgin}.
    \item Propagation-based methods outperform embedding methods on most datasets, indicating the importance of modeling long-range connectivities. We attribute their success to the information aggregation schemes. Specifically, recursively collecting information from neighboring nodes is able to capture the high-order latent semantics for high-quality representation.
    \item Hyperbolic-based methods (\ie Hyper-Know and LKGR) achieve competitive performance compared with Euclidean methods. However, they still slightly underperform state-of-the-art propagation methods (\ie KGIN). This is because they fail to fully exploit the expressive power of hyperbolic space for comprehensive representations. Specifically, Hyper-Know embeds KG in hyperbolic space and directly optimizes the entities' embeddings with triplets, without taking the high-order knowledge associations into consideration. LKGR designs an Euclidean attention mechanism for hyperbolic entities' aggregation, which cannot preserve the local similarities of embeddings in hyperbolic space.
\end{itemize}

\begin{table}[t]
    \caption{Overall performance comparison.}
    \centering
    \vspace{-10pt}
    \label{tab:overall-performance}
    \resizebox{0.48\textwidth}{!}{
    \begin{tabular}{c|c c |c c| c c}
    \hline
    \multicolumn{1}{c|}{\multirow{2}*{}}&
    \multicolumn{2}{c|}{Alibaba-iFashion} &
    \multicolumn{2}{c|}{Yelp2018} &
    \multicolumn{2}{c}{Last-FM} \\
      &recall & ndcg & recall & ndcg & recall & ndcg\\
    \hline
    \hline
    MF  & 0.1095 & 0.0670 & 0.0627 & 0.0413 & 0.0724 & 0.0617 \\
    CKE & 0.1103 & 0.0676 & 0.0653 & 0.0423 & 0.0732 & 0.0630 \\
    UGRec & 0.1006 & 0.0621 & 0.0651 & 0.0419 & 0.0730 & 0.0624 \\
    \hline
    KGNN-LS & 0.1039 & 0.0557 & 0.0671 & 0.0422 & 0.0880 & 0.0642 \\
    KGAT & 0.1030 & 0.0627 & \underline{0.0705} & \underline{0.0463} & 0.0873 & 0.0744 \\
    CKAN & 0.0970 & 0.0509 & 0.0646 & 0.0441 & 0.0812 & 0.0660 \\
    KGIN & \underline{0.1147} & \underline{0.0716} & 0.0698 & 0.0451 & \underline{0.0978} & \underline{0.0848} \\
    \hline
    Hyper-know  & 0.1057 & 0.0648 & 0.0685 & 0.0447 & 0.0948 & 0.0812 \\
    LKGR & 0.1033 & 0.0612 & 0.0679 & 0.0438 & 0.0883 & 0.0675\\
    \hline
    HAKG & \textbf{0.1319$^*$} & \textbf{0.0848$^*$} & \textbf{0.0778$^*$} & \textbf{0.0501$^*$} & \textbf{0.1008$^*$} & \textbf{0.0931$^*$} \\
    \hline
    \hline
    \%Imp. & 14.99\% & 15.43\% & 10.35\% & 8.21\% & 3.07\% & 9.79\% \\
    \hline
    \end{tabular}}
    \vspace{-10pt}
\end{table}

\subsection{Study of HAKG (RQ2)}
\label{sec:ablation_study}
As the hierarchical modeling is at the core of HAKG, we also conduct ablation studies to investigate the effectiveness. Specifically, how the presence of angle loss and gated aggregation, the hyperbolic embeddings and hierarchical relations, the dual item embeddings, and the number of propagation layers affect our model.

\subsubsection{\textbf{Impact of Angle Loss \& Gated Aggregation.}}
\label{sec:exp_ablation}

We first verify the effectiveness of the angle loss and gated aggregation. To this end, three variants of HAKG are constructed by (1) discarding the angle constraint and gated aggregation scheme, termed as HAKG$_{\text{w/o A\&G}}$, (2) removing the angle loss for hierarhical triplets, called HAKG$_{\text{w/o A}}$, and (3) replacing the gated aggregation with simple point-wise addition, named HAKG$_{\text{w/o G}}$. We summarize the results in Table~\ref{tab:impact-of-angle-loss-gate}.

\begin{table}[t]
    \caption{Impact of angle loss and gated aggregation.}
    \centering
    \vspace{-10pt}
    \label{tab:impact-of-angle-loss-gate}
    \resizebox{0.465\textwidth}{!}{
    \begin{tabular}{l|c c |c c| c c}
    \hline
    \multicolumn{1}{c|}{\multirow{2}*{}}&
    \multicolumn{2}{c|}{Alibaba-iFashion} &
    \multicolumn{2}{c|}{Yelp2018} &
    \multicolumn{2}{c}{Last-FM} \\
      &recall & ndcg & recall & ndcg & recall & ndcg\\
    \hline
    \hline
    w/o A\&G  & 0.1218& 0.0799& 0.0737& 0.0458& 0.0946& 0.0872\\
    w/o A & 0.1272& 0.0825& 0.0763& 0.0485& 0.0963& 0.0907\\
    w/o G & 0.1253& 0.0817 & 0.0758& 0.0471& 0.0959& 0.0894\\
    \hline
    \end{tabular}}
    \vspace{-5pt}
\end{table}

\begin{table}[t]
    \caption{Impact of hierarchical modeling.}
    \centering
    \vspace{-10pt}
    \label{tab:impact-of-hierarchical}
    \resizebox{0.465\textwidth}{!}{
    \begin{tabular}{l|c c |c c| c c}
    \hline
    \multicolumn{1}{c|}{\multirow{2}*{}}&
    \multicolumn{2}{c|}{Alibaba-iFashion} &
    \multicolumn{2}{c|}{Yelp2018} &
    \multicolumn{2}{c}{Last-FM} \\
      &recall & ndcg & recall & ndcg & recall & ndcg\\
    \hline
    \hline
    Euclidean  & 0.1231 & 0.0798 & 0.0756 & 0.0484 & 0.0981 & 0.0916 \\
    PH-Relation & 0.1317& 0.0845& 0.0772& 0.0494& 0.1001& 0.0928\\
    GH-Relation & 0.1320& 0.0845& 0.0776& 0.0498& 0.1005& 0.0929\\
    \hline
    \end{tabular}}
    \vspace{-10pt}
\end{table}

Compared with the complete model of HAKG in Table~\ref{tab:overall-performance}, the absence of the angle constraint and gated aggregation dramatically degrades the performance, indicating the necessity of modeling hierarchical relation and collaborative signals. Specifically, HAKG$_{\text{w/o A\&G}}$ directly fuses the dual item embeddings for user aggregation, and ignores the hierarchical relations in KG, and thus, it fails to profile item properly and propagate comprehensive information for learning use. Analogously, leaving the hierarchical relations unexplored (\ie HAKG$_{\text{w/o A}}$) also downgrades the performance. Although HAKG$_{\text{w/o G}}$ retains the modeling of hierarchical relations for characterizing items, it is unable to provide discriminative signals for identifying user behavior patterns, incurring suboptimal user representations.

\subsubsection{\textbf{Impact of Hierarchical Modeling.}}
\label{sec:exp_hierarchiy}
We then evaluate the influence of hierarchical modeling by considering both hierarchical structure and relations. To be more specific, we propose three alternative models, which are modified by: i) replacing all hyperbolic operations with Euclidean alternatives and retaining the computation logic of HAKG (called Euclidean), ii) predicting the hierarchical types of relations by widely used Krackhardt scores criteria~\cite{krackhardt2014graph,nips21conE} in order to compensate for the unavailability of relation types (named PH-Relation), and iii) leveraging the given hierarchical types of relations  to characterize items (termed as GH-Relation). The results of three alternative models are listed in Table~\ref{tab:impact-of-hierarchical}. We observe that:
\begin{itemize}[leftmargin=*]
    \item The performance degrades for all three datasets when we remove the hyperbolic geometry for HAKG, meaning that modeling user-item interactions and KG in hyperbolic space could enhance the model's expressive power and yield better representations for recommendation.
    \item When the hierarchical types of KG relations are not explicitly available, the performance of alternative approaches (\ie PH-Relation) approximates to that of model with ground-truth hierarchical relations (\ie GH-Relation), which empirically shows that our model is robust for different datatsets even when the KG's hierarchical information is not available.
\end{itemize}

\begin{table}[t]
    \caption{Impact of dual item embedding.}
    \centering
    \vspace{-10pt}
    \label{tab:dual_item_emb}
    \resizebox{0.455\textwidth}{!}{
    \begin{tabular}{l|c c |c c| c c}
    \hline
    \multicolumn{1}{c|}{\multirow{2}*{}}&
    \multicolumn{2}{c|}{Alibaba-iFashion} &
    \multicolumn{2}{c|}{Yelp2018} &
    \multicolumn{2}{c}{Last-FM} \\
      &recall & ndcg & recall & ndcg & recall & ndcg\\
    \hline
    \hline
    Single  & 0.1186 & 0.0755 & 0.0769 & 0.0492 & 0.0989 & 0.0904 \\
    Dual  & 0.1319 & 0.0847 & 0.0778 & 0.0501 & 0.1008 & 0.0931 \\
    \hline
    \end{tabular}}
    \vspace{-5pt}
\end{table}

\begin{table}[t]
    \caption{Impact of the number of layers $L$.}
    \centering
    \vspace{-10pt}
    \label{tab:impact-of-layer-number}
    \resizebox{0.465\textwidth}{!}{
    \begin{tabular}{l|c c |c c| c c}
    \hline
    \multicolumn{1}{c|}{\multirow{2}*{}}&
    \multicolumn{2}{c|}{Alibaba-iFashion} &
    \multicolumn{2}{c|}{Yelp2018} &
    \multicolumn{2}{c}{Last-FM} \\
      &recall & ndcg & recall & ndcg & recall & ndcg\\
    \hline
    \hline
    HAKG-1  & 0.1313 & 0.0845 & 0.0766 & 0.0489 & 0.0972 & 0.0897 \\
    HAKG-2  & 0.1306 & 0.0831 & 0.0778 & 0.0501 & 0.0988 & 0.0913 \\
    HAKG-3 & 0.1319 & 0.0848 & 0.0774 & 0.0498 & 0.1008 & 0.0931 \\
    \hline
    \end{tabular}}
    \vspace{-10pt}
\end{table}


\subsubsection{\textbf{Impact of Dual Item Embeddings.}}
\label{sec:exp-dual-item-embeddings}

To analyze the effectiveness of the design of dual item embeddings, we compare it with the conventional single item embeddings, which removes the additional collaborative item embeddings and directly leverages the knowledge embeddings of items in Eq.~(\ref{equ:kg-final-aggregation}) for users' aggregation. We compare their performance on three datasets, and present the results in Table~\ref{tab:dual_item_emb}. We have the following findings:
\begin{itemize}[leftmargin=*]
    \item Discarding the collaborative item embeddings would consistently degrade the performance cross three datasets. This is because if we model items with single representations, they inevitably suffer from information loss when performing the neighbor aggregation. In other words, since items are inherently exhibit two kinds of semantics (collaboration and knowledge), it is unlikely to simultaneously incorporate all the high-order information of items with single embeddings.
    \item The performance of single item embeddings decrease more dramatically on Aliabab-iFashion dataset, compared with it on other two datasets. One possible reason is that the amount of auxiliary information provided by KG is the smallest, which makes it more important to capture the collaborative signals of items with additional collaborative item embeddings.
\end{itemize}

\subsubsection{\textbf{Impact of Model Depth.}}
\label{sec:exp-model-depth}
We also explore the impact of the number of aggregation layers. Stacking more layers is able to collect the high-order collaborative signals and knowledge associations for better capturing of the latent user behavior patterns but at a higher cost. Here, we search $L$ in the range of $\{1,2,3\}$, and report the results in Table~\ref{tab:impact-of-layer-number}. We have the following observations:
\begin{itemize}[leftmargin=*]
    \item Generally speaking, increasing the aggregation layers can enhance the performance, especially for Last-FM datasets. We attribute such improvement to two reasons: (1) Gathering more relevant collaborative signals and knowledge association could provide informative semantics for learning high-quality representations, deepening the understanding of user interest. (2) The dual item embeddings explicitly encode both items' behavior and attribute similarities, which profile item at a finer-grained way than single item embeddings used in other methods.
    \item It is worth mentioning that our model is less sensitive to the model depth, compared with other propagation-based methods~\cite{kdd19kgat,sigir20ckan,www21kgin}. Specifically, HAKG could achieve competitive performance even when $L=1$.
    This is because our separate aggregation schemes and gate mechanism can directly capture useful patterns from both user-item interactions and KG, while the KG-oriented aggregation schemes need more layers to encode the latent semantics from the mixed and obscure information.
\end{itemize}

\begin{figure}[t]
	\centering
	\vspace{-2pt}
	\includegraphics[width=0.21\textwidth]{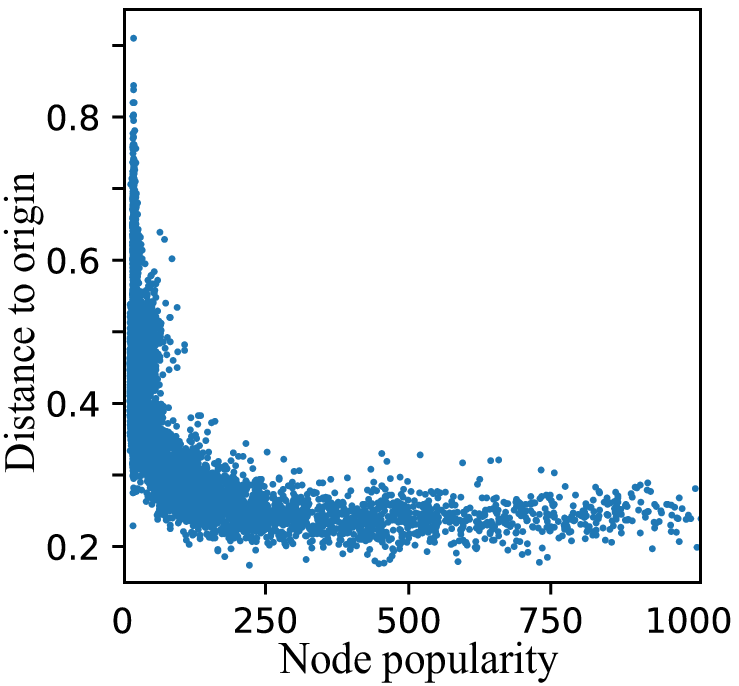}
	\hspace{0.7mm}
	\includegraphics[width=0.24\textwidth]{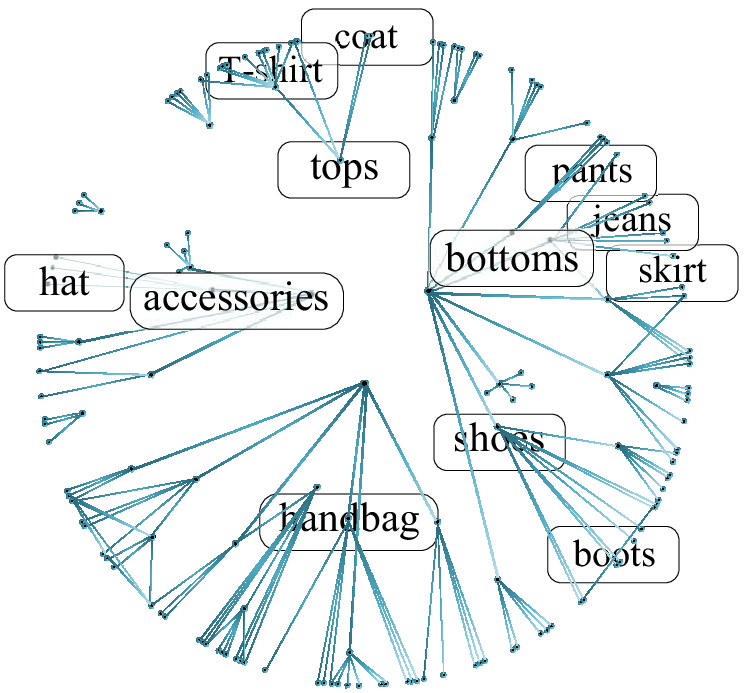}
	\vspace{-2mm}
	\caption{(a) Distance from node (user or item) embedding to origin vs node popularity. (b) Hyperbolic embeddings of entities in KG.}
	\label{fig:vis_kg}
	\vspace{-10pt}
\end{figure}

\subsection{Hierarchies Visualization (RQ3)}
\label{sec:visualization}
In this section, we visualize the learned embeddings to give an intuitive impression of our hierarchical modeling. Specifically, we first train HAKG with two-dimensional embeddings on the Alibaba-iFashion dataset, and separately analyze the hierarchies that exhibit in $\Set{G}_b$ and $\Set{G}_k$:
(1) For the underlying hierarchical structure in interactions, we plot the distance from the origin to hyperbolic embeddings (\eg user embeddings or collaborative item embeddings) versus its popularity. (2) For the hierarchical relations in KG, we select representative entities and demonstrate their connectivities with triplets. As shown in Figure~\ref{fig:vis_kg}, we find that:
\begin{itemize}[leftmargin=*]
    \item The left of Figure~\ref{fig:vis_kg} indicates a clear exponential trend that distance to the origin increases exponentially for less popular items, which is consistent with the degree distribution of interactions in Figure~\ref{fig:distribution_alibaba}. This confirms that our model takes advantage of the exponentially growing volume in hyperbolic space, and uses it to naturally represent the users and items.
    \item The connectivities in the right of Figure~\ref{fig:vis_kg} show clear hierarchical relations between entities. For instance, the entity \textit{bottoms} is categorized into \textit{pants}, \textit{jeans}, and \textit{skirt}, and the attribute information of items (\eg \textit{T-shirt} is a type of \textit{tops}) can be naturally preserved with the hierarchical relations. Thus, HAKG is able to capture the items' attribute semantics in the embedding space, and yield better item representations for learning user preferences.
\end{itemize}

\section{Related Work}
\label{sec:related_work}

Existing recommendation systems incorporated with KG information can be mainly categorized into three clusters, viz., \emph{embedding-based methods}, \emph{path-based methods}, and \emph{propagation-based methods}. We briefly review them in the following.
\begin{itemize}[leftmargin=*]
    \item \textbf{Embedding-based methods}~\cite{kdd16cke,www18dkn,alg18cfkg,sigir18seqkg,www19ktup,sigir20chorus,sigir21UGRec} directly embed entities and relations in KG via knowledge graph embedding (KGE) methods (\eg TransR~\cite{aaai15transR} and TransD~\cite{acl15transD}) to serve as item embedding in recommendation. For example, CKE~\cite{kdd16cke} utilizes TransR to learn item structural representations from knowledge graph, and feeds the learned embeddings to matrix factorization (MF)~\cite{uai09BPRloss} in an integrated framework. Hyper-Know~\cite{aaai21hyperKGRec} embeds knowledge graph in Poincar{\'e} Ball and then designs an adaptive regularization mechanism to regularize item representations. Although these methods benefit from the simplicity and expressiveness of KGE, they fail to capture high-order dependence of user-item relations for user preference learning.
    \item \textbf{Path-based methods}~\cite{recsys16path,kdd18metapath,recsys18rkge,aaai19explainable,www19jointlykg,kdd20hinrec} aim to find semantic paths in KG, and then connect items and users to discover long-range connectivity for recommendation. Those paths can be used to predict user preference with recurrent neural networks~\cite{aaai19explainable,recsys18rkge} or attention mechanism~\cite{kdd18metapath}. For instance, KPRN~\cite{aaai19explainable} captures the sequential dependence within a knowledge-aware path to infer the underlying high-order relation of a user-item interaction. However, defining proper meta-path patterns requires domain knowledge, which can be 
    extremely time-consuming for complicated KG with various types of entities and relations. Moreover, domain-specific meta-paths inevitably lead to poor generalization for different  recommendation scenarios~\cite{kdd20hinrec,kdd20metahin}.
    \item \textbf{Propagation-based methods}~\cite{cikm18ripple,kdd19kgat,kdd19kgnn-ls,www19kgcn,sigir20ckan,www21kgin,cikm21kcan} are inspired by the recent advances of graph neural networks (GNNs)~\cite{iclr17gcn,nips17sageGCN,iclr18gat,icml19SGCN,sigir19ngcf,sigir20lighGCN}, which iteratively perform information aggregation mechanism from neighborhood nodes. As such, these methods are able to discover high-order relations in an end-to-end fashion. For example, KGAT~\cite{kdd19kgat} creates unified knowledge graph (UKG) to combine user-item interactions and KG, and then performs knowledge-aware attention on it. CKAN~\cite{sigir20ckan} employs the heterogeneous propagation strategy to encode the user preference with knowledge associations to further improve the performance. Most recently, KGIN~\cite{www21kgin} considers a new aggregation mechanism to integrate long-range relation paths, and disentangles user preference with intents for better interpretability.
\end{itemize}

\vspace{-2pt}
How to leverage hyperbolic geometry and hyperbolic embeddings in recommender systems has become a recent trend~\cite{wsdm20hyperRs,aaai21hyperKGRec,www21hgcf,kdd21hyperEmbedding,wsdm22hyperRS}. For instance, HyperML~\cite{wsdm20hyperRs} is the first work to leverage hyperbolic geometry for recommendation through metric learning approach; and HGCF~\cite{www21hgcf} proposes a hyperbolic GCN model for collaborative filtering. However, to the best of our knowledge, none of existing methods considers modeling KG relations at a finer-grained level of hierarchies. Moreover, they fail to preserve the crucial high-order collaborative signals of items. Our work differs from them in hierarchical modeling and aggregation, i.e., we aim at addressing the importance of hierarchical relations for profiling items, and explicitly propagate two kinds of information with dual item embeddings for better identifying user behaviors.

\section{Conclusion and Future Work}
\label{sec:conclusions}

In this paper, we propose a new knowledge-aware recommendation model, called HAKG, which captures the underlying hierarchical structure of data in hyperbolic space, and characterize items with hierarchical relations in KG.
Besides, HAKG employs dual item embeddings to separately encode items' collaborative signals and knowledge associations, and develops a gated mechanism to control discriminative signals towards the users' behavior patterns. Extensive experiments conducted on three real-world datasets demonstrate the superiority of HAKG. In the future, we plan to investigate the problem of numerical errors in hyperbolic space~\cite{nips21MCF}, and examine all operations (\eg addition and multiplication) in hyperbolic space to fully exploit the power of hyperbolic geometry. 



\begin{acks}
This work was supported by the NSFC under Grants No. (62025206, 61972338, and 62102351). 
\end{acks}

\bibliographystyle{ACM-Reference-Format}
\balance

\bibliography{ref}
\balance

\end{document}